\documentclass[twocol]{ametsoc}

\usepackage{tabularx}
\usepackage{algorithmic}
\usepackage{algorithm}
\usepackage{amsthm}

\journal{jas}

\bibpunct{(}{)}{;}{a}{}{,}

\title{Cloud Tomography from Space using MISR and MODIS:\\Locating the ``Veiled Core'' in Opaque Convective Clouds}

\authors{Linda Forster
\correspondingauthor{Ludwig-Maximilians-Universit{\"a}t, Theresienstr. 37, 80333 Munich, Germany.}
\thanks{Current affiliation: Jet Propulsion Laboratory, California Institute of Technology, Pasadena, California, USA.}
}

\affiliation{Ludwig-Maximilians-Universit{\"a}t, Munich, Germany
}
\email{linda.forster@physik.lmu.de}

\extraauthor{Anthony B. Davis}
\extraaffil{Jet Propulsion Laboratory, California Institute of Technology, Pasadena, California, USA}

\extraauthor{David J. Diner}
\extraaffil{Jet Propulsion Laboratory, California Institute of Technology, Pasadena, California, USA}

\extraauthor{Bernhard Mayer}
\extraaffil{Ludwig-Maximilians-Universit{\"a}t, Munich, Germany}

\abstract{
For passive satellite imagers, current retrievals of cloud optical thickness and effective particle size fail for convective clouds with 3D morphology.
Indeed, being based on 1D radiative transfer (RT) theory, they work well only for horizontally homogeneous clouds.
A promising approach for treating clouds as fully 3D objects is cloud tomography, and this has been demonstrated for airborne observations.
For cloud tomography from space, however, more efficient forward 3D RT solvers are required.
Here, we present a path forward, acknowledging that optically thick clouds have ``veiled cores.''
Photons scattered into and out of this deep region do not contribute significant information to the observed imagery about the inner structure of the cloud.
We investigate the location of the veiled core for the MISR and MODIS imagers.
While MISR provides multi-angle imagery in the visible and near-IR, MODIS includes channels in the short-wave IR, albeit at a single view angle.
This combination will enable future 3D retrievals to disentangle the cloud's effective particle size and optical thickness.
We find that, in practice, the veiled core is located at an optical distance of $\approx 5$, starting from the cloud boundary along the line-of-sight.
For MODIS' absorbing wavelengths the veiled core covers a larger volume, starting at smaller optical distances.
This result makes it possible to reduce the number of unknowns for the cloud tomographic reconstruction, and opens up new ways to increase the efficiency of the 3D RT solver at the heart of the reconstruction algorithm.
}

\begin{document}

\maketitle

\section{Context, Motivation \& Outline}
\label{sec:intro}

Convective clouds have an important impact on the Earth's weather and climate.
By reflecting solar radiation and trapping thermal radiation, they play a key role in the Earth's radiation budget.
In addition, deep convective clouds are one of the key drivers of the hydrological cycle by redistributing water throughout the depth of the troposphere \citep{jensen2016}.
Shallow convection, especially in trade-wind cumuli, have benn shown to be very sensitive to changes in their environmental conditions with a critical impact on future global warming \citep{bony2017}.
The IPCC AR5 \citep{boucher2013} reports that clouds and aerosols still contribute the largest uncertainty to estimates of the Earth's changing energy budget.
Investigating and monitoring cloud properties is therefore of great importance in order to improve parameterizations of convection in numerical models, for operational weather forecasting as well as for the global climate.
This, in turn, is an essential step towards a better understanding of convective processes as well as aerosol--cloud interactions and their impact on climate.

Quantification and monitoring of cloud properties with global coverage is operationally performed by space-borne passive and active remote sensing.
While active remote sensing (e.g., mm-wave radar) provides information about vertical cloud structure, the advantage of passive remote sensing (e.g., multi-spectral imaging) is the significantly larger spatial coverage.
Traditional passive retrieval methods for cloud optical properties such as cloud droplet effective radius and optical thickness often make use of a dual-wavelength approach using passive reflectance measurements in the visible and near-infrared (VNIR) and short-wave infrared (SWIR) as first developed by \cite{nakajima90}.
While the absorbing wavelength in the SWIR provides sensitivity to particle size, the scattering-dominated VNIR spectral band is more sensitive to cloud optical thickness.
Based on one-dimensional (1D) radiative transfer (RT), this method applies reasonably well to extended stratiform cloud layers far away from cloud edges (e.g., \cite{platnick2003}, \cite{zhang2011}) but is not able to account for three-dimensional (3D) cloud morphologies driven by shallow or deep convection.
This leaves a huge gap in the global observation of cloud optical properties (e.g., \citet{cho2015}).
To close this gap, a retrieval method is needed that is tailored to the 3D morphology of convective clouds and applicable to satellite observations.

3D remote sensing of convective clouds using multi-angle satellite imaging from MISR was pioneered by \cite{seiz2006} using sophisticated stereographic methods borrowed from photogrammetry.
Only the outer shape of the cloud was targeted and, in view of the labor involved, only one massive convective cloud was investigated, and only one 2D transect of the surface was retrieved.
This procedure is entirely based on finding ``features'' that appear in two or more MISR images, which does not call for calibrated radiance fields.
Given this best estimate of the outer cloud boundary in the $(y,z)$ plane, \cite{cornet2008} assumed simple homogeneous or two-layered representations of the extinction coefficient and phase function inside the cloud, with uniformity in the $x$ direction, and then performed Monte Carlo simulations to predict the calibrated MISR radiances.
By trial and error, they found a reasonable match with the observed radiances, and thus determined the optical thickness of this vertically-developed convective cloud.
The same cloud as investigated in detail by Seiz, Cornet and Davies was recently revisited by \cite{lee2018} who used ``space carving'' to determine the an outer bound of the 3D volume occupied by the cloud.
This volume is defined on a 3D cartesian grid based on MISR pixel footprints by intersecting back-projected cloud masks for all nine angles.
Their outcome compared well with the feature-based surface reconstruction by \citeauthor{seiz2006} along the specific transect they focused on.

\citet{levis2015} recently demonstrated full 3D tomographic cloud reconstruction using airborne multi-angle images from JPL's Airborne Multiangle SpectroPolarimetric Imager (AirMSPI, \cite{diner2013}).
Specifically, the authors used reflectance measurements from AirMSPI's red channel for nine different viewing angles and reconstructed the 3D cloud extinction field on a cartesian grid.
\citeauthor{levis2015} treat the reconstruction as a large inverse problem, using as the forward model their customized version of \citeauthor{evans98}' \citeyear{evans98}  popular open-source 3D RT solver, the Spherical Harmonics Discrete Ordinate Method (SHDOM).  
This early cloud tomography demonstration used radiance fields from both synthetic clouds from a Large-Eddy Simulation (LES) model \citep{MatheouChung14} and a real counterpart observed with AirMSPI, where the former provided the truth for retrieval error quantification.  
In both cases, the cloud microphysics were prescribed and, for simplicity, held uniform across the cloud mass, leaving only the large 3D grid of extinction values to be determined.

\citet{levis2017} built on their \citeyear{levis2015} success by extending the retrieval to all AirMSPI channels, but \emph{without} polarization.
Interestingly, even without polarization nor SWIR spectral channels, they gained enough sensitivity to reconstruct the cloud while treating the microphysical parameters (effective radius and variance) as unknown but constrained to vary only along the vertical direction, which is as expected in nature at least in the bulk of the cloud.
Moreover, the reconstruction of the 3D extinction grid becomes more accurate than with the prescribed microphysics.

In order to provide 3D cloud macro- and microphysical properties with global coverage, this computed tomography (CT) method has to be advanced so that it is applicable to satellite observations.
To explore this, we use Terra's Multi-angle Imaging SpectroRadiometer (MISR, \citet{diner1998}) for its multi-angle viewing capability and MODerate-resolution Imaging Spectro-radiometer (MODIS, \citet{king2003}) for its multi-spectral coverage in the SWIR region.
However, transition from the 10~m AirMSPI pixels to larger counterparts of 275~m in the case of MISR (250 and 500~m for MODIS) poses two significant problems: 
first, optically thick pixel-scale volumes will occur and, second, unresolved sub-pixel spatial variability of cloud extinction and microphysics will be present.
SHDOM, the RT model at the heart of the existing 3D reconstruction algorithm is ill-suited for opaque clouds.
Optically thick pixels will be subdivided by SHDOM's adaptive grid refinement into very many optically thinner cells, causing it to rapidly swamp computer memory.
Moreover, there is no prior information about the specific kind of unresolved variability, yet, neglecting it invariably leads to biased results.
To overcome these issues, a new approach must be developed to perform RT inside optically thick clouds more efficiently.

In the present study, we examine the hypothesis that optically thick clouds contain a ``veiled'' core.
Inside this core, details of the liquid water distribution do not significantly contribute to satellite observations.
These details are thus ``veiled'' by multiple scattering throughout the cloud volume.
The 3D reconstruction algorithm makes use of the information about the cloud optical properties carried by the detected photons.
This intricate spatial information originates primarily from the outer shell of the cloud, whereas the amount of information conveyed about the inner core of the cloud decreases significantly with depth due to multiple scattering. 
Being able to separate the cloud's ``veiled'' core from its shell will facilitate new, more efficient ways of solving the 3D RT inside potentially very large opaque clouds.

In this paper we investigate if a veiled core exists and where it is located inside optically thick clouds from a space-borne perspective using synthetic MISR and MODIS observations.
MISR provides the key multi-angle information in the VNIR.
SWIR observations are also required for the new approach of 3D cloud tomography from space to disentangle cloud optical thickness and droplet effective radius \citep{nakajima90}, even if they are at a single view angle and at a degraded spatial resolution.
This will be achieved by fusing data from MISR and MODIS.

The paper is organized as follows.
In Section~\ref{sec:methods}, we describe the stochastic model we used to represent vertically-developed convective clouds in which we seek the veiled core under a range of opacity and illumination conditions; full details are provided in Appendix~B.  
We also describe the adopted numerical 3D RT modeling framework.
In Section~\ref{sec:resultsMISR}, we describe our observational definition of the veiled core, and present our findings for the adopted class of cloud models taking MISR's VNIR perspective; further results are presented in Appendix~A for oblique illumination.  
In Section~\ref{sec:resultsMODIS}, we switch from MISR's to MODIS' perspective on the veiled core, with an emphasis on the impact of droplet absorption, hence sensitivity to particle size, in SWIR channels.
We summarize our findings in Section~\ref{sec:concl} and describe future research in support of cloud CT.
In a companion paper \citep{Davis_etal19}, we relate our findings to the fundamentals of RT in opaque optical media.

\section{Methods}
\label{sec:methods}

\begin{table*}[t]
\caption{MISR camera labels and corresponding VZAs.
``n'' denotes nadir, ``a'' aftward and ``f'' forward view.}
\label{tab:misr_cameras}
\begin{center}
\begin{tabular}{c c c c c c c c c}
\hline\hline
Df & Cf & Bf & Af & An & Aa & Ba & Ca & Da\\
\hline
$+70.5^\circ$ & $+60.0^\circ$ & $+45.6^\circ$ & $+26.1^\circ$ & $0^\circ$ & $-26.1^\circ$ & $-45.6^\circ$ & $-60.0^\circ$ & $-70.5^\circ$ \\
\hline\hline
\end{tabular}
\end{center}
\end{table*}

To investigate the veiled core inside optically thick convective clouds from a satellite perspective, RT simulations are performed using synthetic clouds.
In this study, we use the 3D Monte Carlo RT solver MYSTIC \citep{mayer2009,buras2011}, which is part of libRadtran \citep{mayer2005,emde2016}.

\begin{figure}[b]
\includegraphics[width=8.3cm]{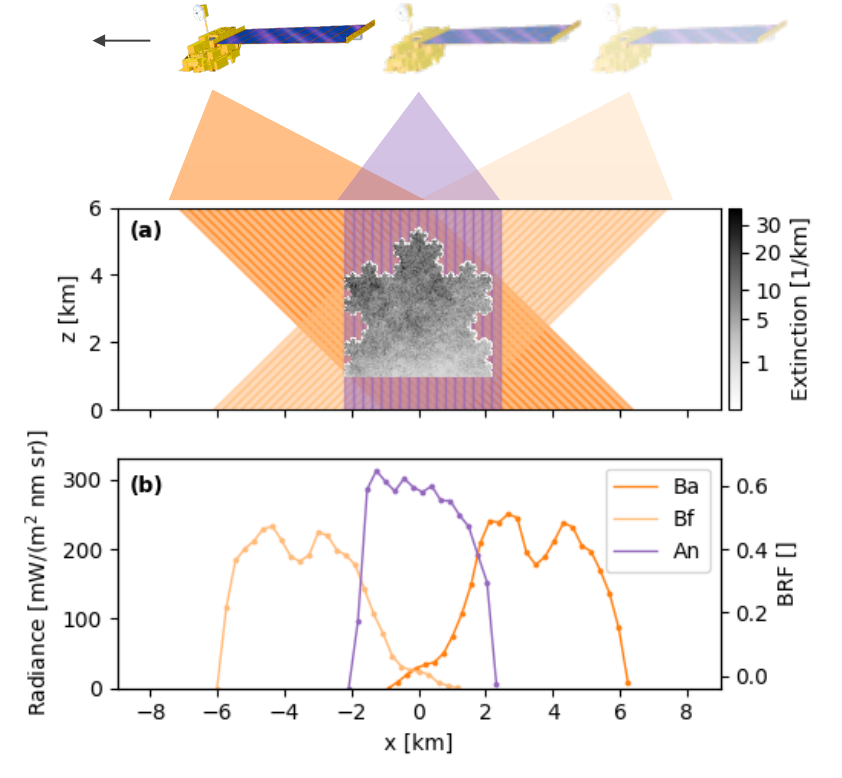}
\caption{
(a) MISR's observation geometry for an overpass over the Koch cloud.
(b) Corresponding ground-registered red-channel (670~nm) radiances $I_\text{cam}(x)$ measured in [mW/m$^2$/nm/sr] (l-h axis) and in [non-dimensional] ``Bidirectional Reflection Function'' BRF = $\pi I_\text{cam}(x)/\mu_0 F_0$ units (r-h axis, with TOA irradiance $F_0$ = 1.5290e3 mW/m$^2$/nm and $\mu_0 = \cos$SZA = 1 in this case).  Three of the nine cameras are displayed, An in purple and the Bf and Ba cameras at $\pm45^\circ$ in (pastel) orange; see Fig.~\ref{fig:koch_highres_orig} for all nine views.
The shaded areas in (a) symbolize the respective viewing angles with each bin corresponding to an along-track (``push-broom'') pixel $[x,x+\Delta x)$ in (b) with a spatial resolution of $\Delta x$ = 275~m.
}
\label{fig:misr_geo}
\end{figure}

Simulations are performed for MISR's red band at 670~nm for each of the nine viewing angles and a spatial resolution of 275~m.
The nine MISR cameras are labeled An, Af/Aa, Bf/Ba, Cf/Ca, and Df/Da \citep{diner1998}.
The labels range from ``A'' for the central cameras to ``D'' for the most oblique cameras, with ``n'' representing nadir, ``a'' aftward and ``f'' forward viewing direction along track.
The corresponding viewing zenith angles (VZA) are listed in Table~\ref{tab:misr_cameras} where a positive (negative) VZA points the camera forward (aftward).
MISR's observation geometry for three of its nine cameras is displayed in Fig.~\ref{fig:misr_geo}.
MISR is passing over the cloud depicted in panel (a) from North to South (right to left in the image) on its descending daytime orbit, as indicated by the arrow.
First, the forward facing cameras (exemplified here by the Bf camera at 45.6$^\circ$ in pastel orange) observe the cloud, followed by the An (purple), and finally the aftward facing cameras (exemplified here by Ba, orange).

RT simulations for MODIS are performed for the channels 1, 5, 6, and 7, assuming central wavelengths of 645, 1240, 1640, and 2130~nm \citep{barnes1998}.
The spatial resolution of MODIS' channels 5 to 7 is 500~m and for channel~1 is 250~m.
For these simulations a single VZA of 9$^\circ$ is chosen as a typical value for a cloud at a random position within MISR's $\approx$400~km cross track swath width, which is near the middle of MODIS' 2330~km \citep{king2003}.
MISR and MODIS are both aboard NASA's Terra spacecraft that was launched in December 1999 in a near-polar and sun-synchronous orbit at an altitude of 705~km and which crosses the equator at 10:30 a.m. on the descending node. 

\begin{figure*}[t]
\includegraphics[width=\textwidth]{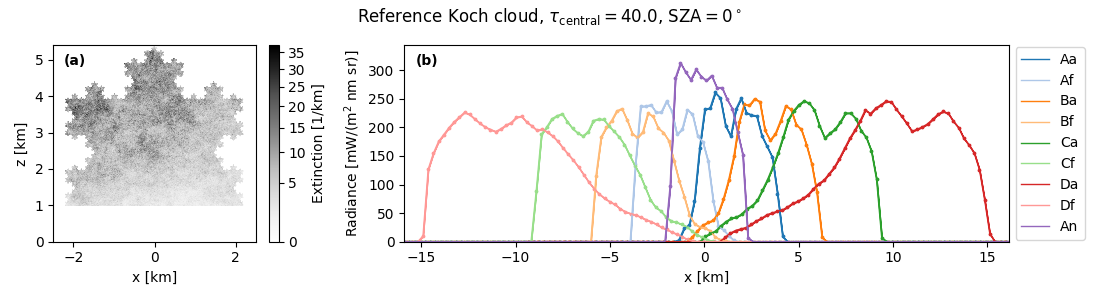}
\caption{
(a) Extinction field of the reference Koch cloud with a central optical thickness of $\tau_\mathrm{central} = 40$.
(b) MISR radiances simulated with MYSTIC for the nine cameras Aa (Af) through Da (Df) and An registered to the ground.
The sun is overhead ($\mathrm{SZA}=0^\circ$) in these 3D RT simulations.
The uncertainty of the results is represented by the width of the lines and amounts to less than 1\% for a 2$\sigma$ confidence interval.
The symbols indicate the exact location of the sampled pixels along the $x$-axis, and are all larger than the 1\% Monte Carlo error.
}
\label{fig:koch_highres_orig}
\end{figure*}

The sensitivity studies presented in the following use idealized 2-dimensional (2D) convective clouds as shown in Fig.~\ref{fig:misr_geo}a and \ref{fig:koch_highres_orig}a.
The synthetic clouds vary only in the $x$- and $z$-direction with internal turbulence simulated by a 2D high-resolution fractional Brownian motion (fBm) field \citep{Mandelbrot77}.
Their outer shape is derived from a fractal Koch curve \citep{koch1904}.
Close examination of the figures shows that the extinction field exhibits a modest vertical gradient in the horizontally averaged value.
This gradient was added to mimic the vertical structure of realistic convective clouds.
Airborne in situ measurements have shown that cloud extinction increases with altitude inside the cloud \citep{pawlowska2000}.
See Appendix~B for more details on the generation of the synthetic clouds.
We chose this simplified setup to reduce computational costs and help interpret the results; in particular, ``images'' in a 2D world can be displayed simply as mono-variate functions of the $x$ coordinate.

Figures~\ref{fig:misr_geo}a and \ref{fig:koch_highres_orig}a show the resulting cloud extinction field scaled to a central optical thickness of $\tau_\mathrm{central}=40$.
This synthetic cloud is used as an input field for the RT solver MYSTIC.
The $1025\times1024$ pixels are embedded in a $\approx$53~km domain with a horizontal and vertical resolution of about 4.3~m, resulting in a cloud width and physical thickness of about 4.4~km with a cloud base height at 1~km, hence a cloud top height of 5.4~km (cf. Fig.~\ref{fig:misr_geo}a).
The domain size is chosen large enough to avoid artifacts caused by periodic boundary conditions, even for the most oblique MISR viewing angles ($\pm$70.5$^\circ$ for Df/Da cameras).  
For MYSTIC's purposes, a space-based sensor is located at the top of the computational grid.
Therefore, to keep the domain size, and thus computational time, as small as possible, the grid top is located just above the cloud at 6~km altitude.

For simplicity, the surface albedo is set to zero (pure absorption) and the solar zenith angle (SZA) is set to 0$^\circ$ (overhead sun).
For all simulations a constant effective droplet radius of $r_\mathrm{eff}=10$~$\mu$m is chosen, with optical properties computed according to Mie theory \citep{wiscombe80a}.
Aerosol as well as molecular scattering and absorption are neglected to focus on the cloud properties only.
Finally, the RT simulations are calibrated with the solar spectrum by \cite{thuillier2003}.

Using this gridded cloud/surface scene, the simulated radiances as observed at the nine MISR cameras are computed at the top of the MYSTIC grid, and subsequently projected to the ground in order to simulate correctly actual MISR observations.
The radiances of the high-resolution cloud field are then averaged over 64 pixels to obtain MISR's resolution of 275~m.
The results of three representative cameras (Bf, An, Ba) are displayed in Fig.~\ref{fig:misr_geo}b, and the full suite of 1D ``images'' are displayed in \ref{fig:koch_highres_orig}b.
The numerical uncertainty was controlled and maintained at a level far below 3\%, MISR's absolute radiometric error at maximum signal \citep{diner1998}.

For the following sensitivity studies, we assume that, for an average signal, differences in the simulated MISR radiances of $\lesssim 5\%$ are too close to the instrument's noise, and thus considered negligible.
In other words, we assume a sensitivity threshold for MISR of 5\%, which corresponds to the root-mean-square (RMS) error on the difference between two independent MISR radiances at the 1.2$\sigma$ level.
To resolve these differences in the RT simulations, we determined that a numerical precision of $\le 1\%$ is required.
This is achieved by performing simulations with MYSTIC using $2 \times 10^5$ photons per cloudy pixel, or $1.28 \times 10^7$ photons per MISR pixel (64 aggregated cloud pixels).

We hypothesize that changes in the liquid water content (LWC) inside the veiled core lead to negligible differences in the observed MISR as well as MODIS radiances in the above sense of being $\lesssim$5\%.
To investigate the location of the veiled core inside these synthetic clouds, the LWC distribution in their inner core is manipulated and the observed radiances are compared to the reference cloud, as explained in the following.

Sensitivity studies are performed with a reference cloud field and three different scenarios of modifying the 2D LWC grid:
in the first scenario, the LWC in the inner core of the reference Koch cloud is replaced by its mean value.
In the other two scenarios, the liquid water distribution in the cloud's inner core is replaced by new realizations of the turbulence field.
To investigate the location of the veiled core inside these synthetic clouds, the simulated radiances of the reference case are compared with the radiances of both cloud variations.
While MISR's red band is controlled entirely by scattering, absorption plays an important role for the MODIS' SWIR channels and its effect is investigated in \ref{sec:resultsMODIS}.
First, we present sensitivity studies for locating the veiled core using MISR observations.

\section{MISR's perspective on the veiled core}
\label{sec:resultsMISR}

The optical thickness of the reference Koch cloud in Fig.~\ref{fig:koch_highres_orig}a is scaled to $\tau_\mathrm{central} = 40$ along the central vertical pixel column.
Figure~\ref{fig:koch_highres_orig}b shows the corresponding MISR radiances computed for the nine cameras described in Table~\ref{tab:misr_cameras}.
The simulated radiances are projected to the ground assuming a satellite overpass from North to South, as shown in Fig.~\ref{fig:misr_geo}.
This projection is precisely how MISR's Level 1 radiances and Level 2 (e.g., stereo-matching) products are archived, which is to register the observations to the WGS84 ellipsoid \citep{diner1998}.
MISR radiances, as observed by the nadir pointing camera (An), are displayed at the center of the domain in purple.
On the right hand side of the An camera, the radiances observed by the aftward pointing cameras are represented in blue (Aa), orange (Ba), green (Ca), and red (Da).
On the left hand side the results of the forward pointing cameras are shown in pastel blue (Af), orange (Bf), green (Cf), and red (Df).
The uncertainty of the MYSTIC simulations is represented here by the width of the lines, which amounts to less than 1\% for a 2$\sigma$ confidence interval.
Assuming an overhead sun with $\mathrm{SZA} = 0^\circ$ causes the detected radiances of the nadir An camera to be brightest.
Since the geometry of the Koch cloud is symmetric around its vertical central axis, asymmetries in the radiances are due to large-scale turbulence in the extinction field.
Darker shades correspond to larger extinction values, which in turn cause larger reflected radiances.
This asymmetry is clearly visible, for instance, for the An camera comparing the left vs. right side of the cloud.
How the spatial footprint of the cloud for the cameras characteristically increases for more oblique angles is also noticeable.

\begin{figure*}[t]
\includegraphics[width=\textwidth]{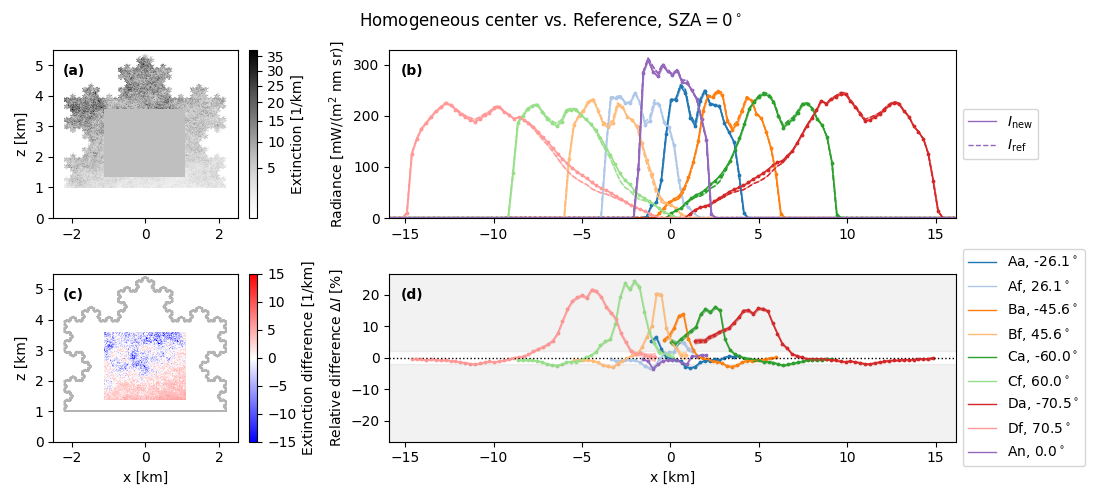}
\caption{
(a) Extinction field of the reference cloud from Fig.~\ref{fig:koch_highres_orig} manipulated by replacing the extinction values of its inner core by their average of 8.34~km$^{-1}$, which results in $\tau_\mathrm{central} \approx 38$.
(b) MISR radiances registered to the ground for the nine cameras showing the results for the reference cloud (dashed) together with the results of the new cloud field (solid).
(c) Absolute difference of the extinction field in [km$^{-1}$], new extinction field minus reference.
The gray curve indicates the outline of the cloud.
(d) Relative differences in [\%] of the observed MISR radiances: new minus reference cloud relative to reference.
The uncertainty of the relative difference is represented by the shaded area around each line, corresponding to a $2\sigma$ confidence interval.
The white area around the black dashed zero-line indicates the $\pm$5\% threshold.
Within this range, differences in the observed radiances are considered below the instrument's noise level and thus negligible, but here most differences far exceed that threshold.
}
\label{fig:koch_highres_homcenter}
\end{figure*}

Figure~\ref{fig:koch_highres_homcenter} shows the first scenario of manipulating the cloud core.
Here, the internal core of the reference Koch cloud is replaced by its mean value.
Size and shape of the core region is chosen arbitrarily in this first approach but aiming at covering a significant portion of the cloud's volume.
The resulting extinction field is displayed in Fig.~\ref{fig:koch_highres_homcenter}a.
Figure~\ref{fig:koch_highres_homcenter}b compares the MISR radiances for the reference ($I_\mathrm{ref}$, dashed line) and the manipulated cloud ($I_\mathrm{new}$, solid line).
The extinction difference (reference minus new) is depicted in Fig.~\ref{fig:koch_highres_homcenter}c.
Furthermore, the relative differences between the observed radiances of the reference cloud and its modified version $\Delta I/I$ are computed according to
\begin{equation}
\frac{\Delta I}{I} = \frac{I_\mathrm{new} - I_\mathrm{ref}}{I_\mathrm{ref}} \cdot 100 \; [\%]
\label{eq:rel_diff}
\end{equation}
and shown in Fig.~\ref{fig:koch_highres_homcenter}d.
The numerical uncertainty of the relative difference is represented by the shaded area around each line, corresponding to a $2\sigma$ confidence interval.
The white area centered around the black dashed zero line, indicates a threshold of $\pm$5\%.
As previously explained, relative differences of simulated radiances within this threshold range are considered negligible in view of the instrument's radiometric noise.

In this sensitivity study, most camera pixels detect a significant difference between the manipulated and the reference cloud up to 26\%.
These relative differences are slightly larger for the forward pointing cameras which observe predominantly the optically thinner part of the cloud field.
The optically thinner part of the cloud allows for larger photon mean free path lengths, i.e., larger penetration depths.
This causes the modified inner region of the cloud to contribute more strongly to the observed radiances compared to the optically thicker cloud area, which is mainly visible for the forward facing cameras.
The nadir together with the Aa and Af views exhibit the overall smallest relative differences with a magnitude up to about 6\%.
Most of the significant relative differences (i.e., $>5\%$) have a positive sign, which implies that the radiances of the manipulated cloud field ($I_\mathrm{new}$) are larger than the reference cloud.
Comparing Fig.~\ref{fig:koch_highres_homcenter}d with the extinction difference in Fig.~\ref{fig:koch_highres_homcenter}c explains these results:
positive relative differences in the MISR radiances correspond to cloud pixels where the extinction was increased by replacing the LWC at the core of the cloud by its mean value.
Increased values in the cloud extinction field are represented by reddish colors in Fig.~\ref{fig:koch_highres_homcenter}c and are mainly detected by the forward facing cameras Bf, Cf, and Df.
Two effects are responsible for the large differences in the observed radiances: first, replacing the turbulent medium by its mean value reduces the length of the effective photon mean-free-path \citep{DavisMarshak04}; second, by averaging over the inner core, the vertical gradient disappears in the manipulated cloud core.
We therefore have to consider both of these effects when applying the concept of veiled cores to real-world clouds.

\begin{figure*}[t]
\includegraphics[width=\textwidth]{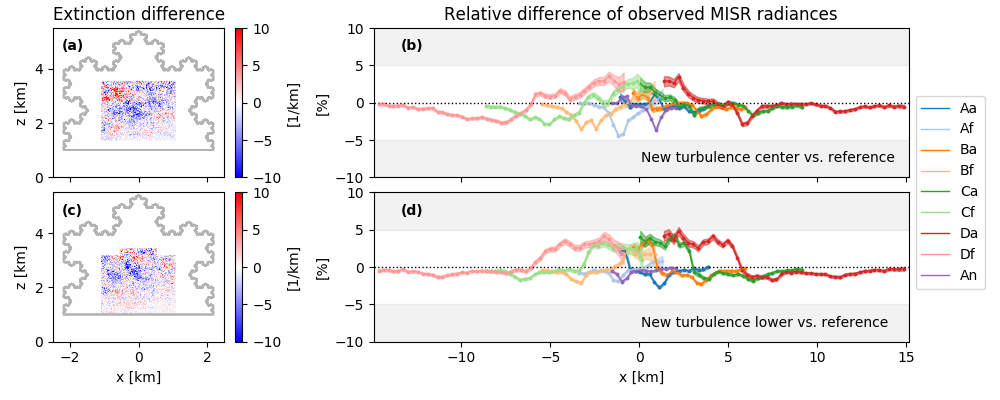}
\caption{
(a) and (c): Absolute difference of the extinction field in [km$^{-1}$] (new minus reference) as in Fig.~\ref{fig:koch_highres_homcenter}c, but for a cloud field generated by replacing the internal extinction core by a new realization of the turbulence field that is constrained to have the same values all along its boundary.
The gray curves indicate the outline of the Koch cloud.
(b) and (d): Relative difference in [\%] of the observed MISR radiances, as in Eq.~(\ref{eq:rel_diff}) and in Fig.~\ref{fig:koch_highres_homcenter}d.}
\label{fig:koch_variations}
\end{figure*}

Figure~\ref{fig:koch_variations} demonstrates similar experiments, this time the core region of the cloud is replaced by a new realization of the internal turbulence field, simulated with fBm.
The same parameters (Hurst index $H$ and, hence, spectral slope $\beta$, cf. Appendix B) are used to model the fBm and a seamless transition to the surrounding cloud volume is ensured using the diamond-square algorithm (cf. Appendix B) .
In this case, the vertical extinction gradient remains the same.
Thus, the new extinction field is \emph{visually} indiscernible from the reference field in Fig.~\ref{fig:koch_highres_orig}a: only the random details in the core region changed.  
This belies the fact that the magnitude of the grid-scale differences at large enough distances from the unchanged core boundary are in fact comparable to those  displayed in Fig.~\ref{fig:koch_highres_homcenter}c where the core extinction is set everywhere to its mean value. 
However, the new simulated MISR radiances (not shown) appear to be the same as in Fig.~\ref{fig:koch_highres_orig}b.
For this reason, cloud extinction and MISR radiances in Fig.~\ref{fig:koch_variations} are only represented by their differences with respect to the reference case.
For Figs.~\ref{fig:koch_variations}a and \ref{fig:koch_variations}b, the central core of the Koch cloud (cf. Fig.~\ref{fig:koch_highres_homcenter}a) was manipulated by regenerating the fBm inside the core.
For Figs.~\ref{fig:koch_variations}c and \ref{fig:koch_variations}d, the liquid water distribution in a lower region of the cloud, which aligns with the cloud bottom at 1~km, was perturbed.
The relative differences $\Delta I/I$ of the simulated radiances are smaller than 5\% for all MISR pixels.
Thus, we can conclude:
\textbf{changing the internal turbulence field results in negligible differences as long as mean, variance, spatial correlations, and any cloud-scale trends are maintained}.

\begin{figure*}[t]
\includegraphics[width=\textwidth]{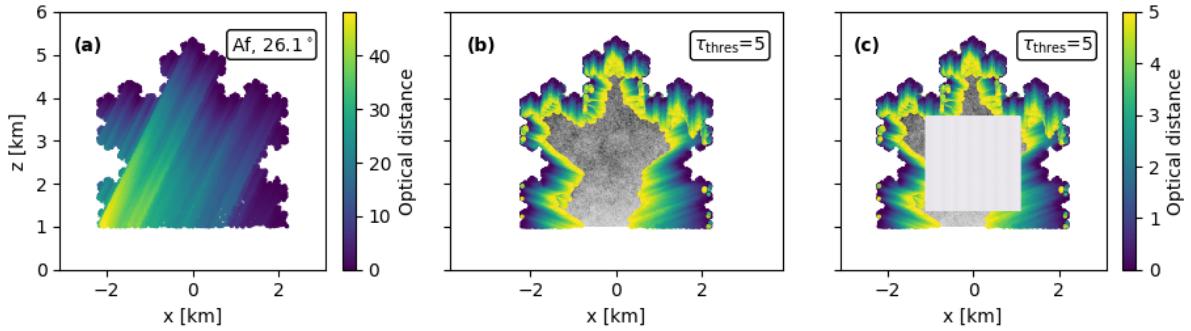}
\caption{
(a) Optical distance along the line of sight of each cell used by MYSTIC at the top of the turbulence grid for the Af MISR camera at a VZA of $-26.1^\circ$ when $\tau_\mathrm{central}$ = 40.
(b) Optical distances for all 9 MISR cameras setting a threshold of $\tau_\mathrm{thres}$ = 5.
The criterion for the masked area is that all pixels of the nine cameras have optical distances larger than $\tau_\mathrm{thres}$.
(c) The veiled core region of the cloud where the turbulence field was modified in Fig.~\ref{fig:koch_variations}a is also displayed.
}
\label{fig:photon_path}
\end{figure*}

The location and size of the veiled core in the previous experiments was decided by no other considerations than maximizing its size (half of the cloud size) and assuming a simple geometric shape at the same time (as required by the specific fBm generation technique we adopted in Appendix B).
To provide a universal method of determining the size and location of the veiled core in arbitrary 3D clouds, a general metric is required, independent of the cloud's geometric shape.
We use the optical distance along the line of sight of each MISR camera, as shown in Fig.~\ref{fig:photon_path}a for the Af camera.
MYSTIC's grid sweep function is used to calculate the optical distance for each MISR camera.
This is achieved in practice by switching off scattering and tracking the photon's location in the $(x,z)$-plane.
To obtain the optical path, the optical thickness at each photon step is integrated along the camera's line-of-sight.
Assuming a certain threshold optical path $\tau_\mathrm{thres}$, the location of the respective cloud region can be determined by applying a mask to the optical distance.
The criterion for the mask is that the optical distance for all nine cameras is larger than the threshold for a certain pixel.
Results for a threshold optical depth of $\tau_\mathrm{thres} = 5$ are displayed in Fig.~\ref{fig:photon_path}b.
Due to the turbluence-induced asymmetry of the cloud extinction field, the forward pointing cameras ``see'' deeper into the cloud, and thus the veiled core is shifted slightly to the left.
Figure~\ref{fig:photon_path}c overlays the region where the turbulence field was modified in the previous section (cf. Fig.~\ref{fig:koch_variations}a).
We can now determine the optical depth from the cloud's outer fractal boundary at which the inner extinction was manipulated for each MISR camera and pixel.

\begin{figure}[t]
\includegraphics[width=8.3cm]{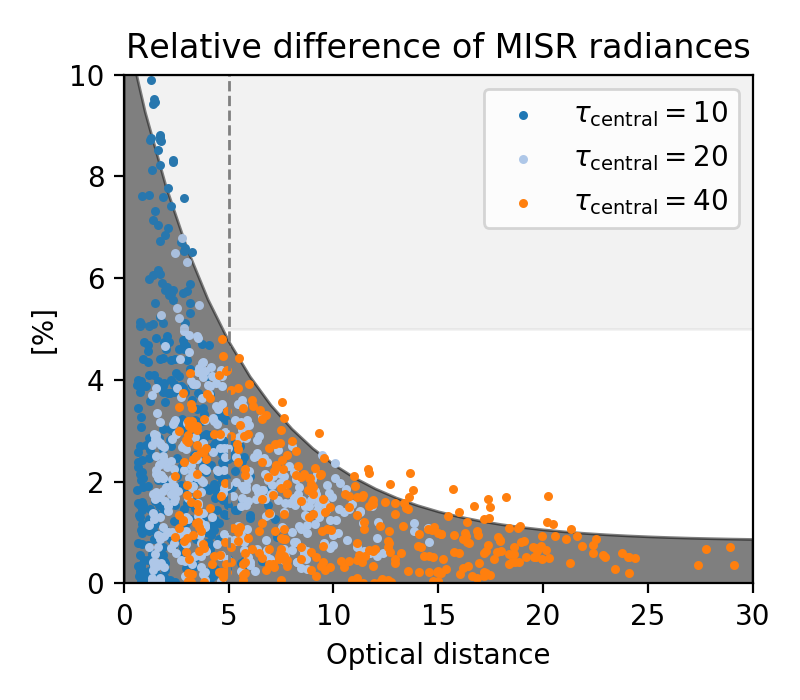}
\caption{
Relative absolute difference of MISR radiances in [\%] of the sensitivity tests shown in Figs.~\ref{fig:koch_variations}b and \ref{fig:koch_variations}d versus the optical distance of the boundary of the veiled core, all MISR cameras considered.
The results were combined for different total cloud optical thicknesses ranging from $\tau_\mathrm{central}$ = 10 (dark blue), 20 (light blue) to 40 (orange).
The vertical dashed black line indicates an optical distance of 5 for the veiled core, which corresponds to a relative difference of less than $\pm$5\%, indicated by the white area.
}
\label{fig:rel_diffs}
\end{figure}

We expect that the relative differences in the observed MISR radiances decrease as the optical distance between cloud surface and manipulated core increases.
In other words, the deeper inside the cloud the manipulation occurs, the more it is veiled by additional multiple scattering.
Figure~\ref{fig:rel_diffs} shows the relative differences of the simulated MISR radiances as a function of increasing optical distance between the cloud surface and the manipulated core.
The relative differences of all nine cameras and all MISR pixels are combined from both simulations shown in Figs.~\ref{fig:koch_variations}b and \ref{fig:koch_variations}d and for different total cloud optical thicknesses ($\tau_\mathrm{central}$ = 10, 20, 40).
This representation yields a simple relationship between the relative difference of the MISR radiances and the optical distance defining the veiled core.
Given a certain sensitivity threshold of the instrument, the location of the veiled core can be directly inferred in terms of optical distance from the cloud surface.

For MISR, we assumed a sensitivity threshold of about $\pm$5\%, which is indicated by the white area in Fig.~\ref{fig:rel_diffs}.
For a veiled core at optical distance 5 (dashed vertical line), all measured relative differences from the three previous experiments shown in Figs.~\ref{fig:koch_variations}b and \ref{fig:koch_variations}d are below 5\%.
This implies that for a sensor-driven uncertainty threshold of 5\%, the veiled core starts at optical distance of $\approx$5 inside the cloud.
Another implication is that, for that particular sensitivity threshold of 5\%, only clouds with total optical thickness in excess of 5 exhibit a veiled core at all.

For larger effective radii, clouds with a fixed LWC have a smaller cloud optical thickness (cf. Eq.~\ref{eq:tau}).
Also, following Mie theory \citep{mie08}, the forward-scattering peak of the phase function narrows slightly.
For a fixed optical thickness, this causes the photons to penetrate slightly deeper, resulting in the veiled core moving deeper inside the cloud.
In this case, the boundary of the threshold optical distance $\tau_\mathrm{thres}$ that defined the veiled core might have to be increased a little.
The opposite is true for smaller effective radii.
This implies that the location of the veiled core in terms of optical distance from the outer cloud boundary is only weakly sensitive to changes in the effective droplet radius as well as to the total cloud optical thickness.
The only limitation is that the total cloud optical thickness must exceed $\tau \ge 5$ for a veiled core to exist.

\section{MODIS' perspective on the veiled core}
\label{sec:resultsMODIS}
%
\begin{figure*}[t]
\includegraphics[width=\textwidth]{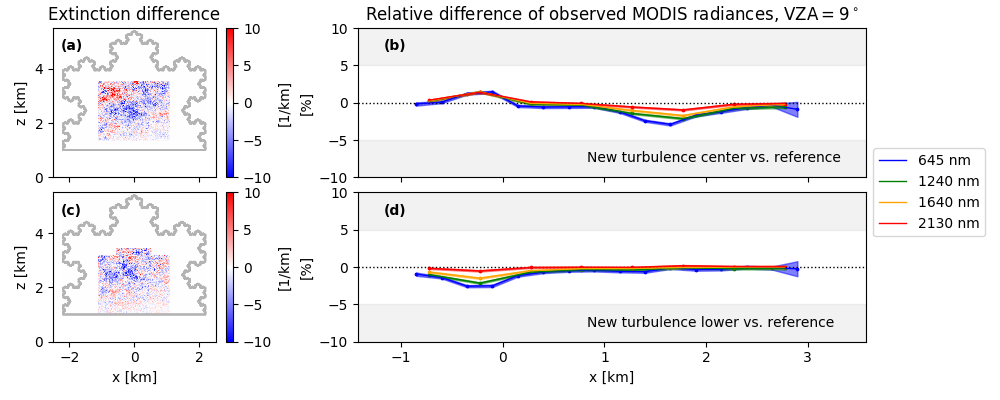}
\caption{
(a) and (c): Absolute difference of the extinction field in [km$^{-1}$] (new vs. reference) as in Fig.~\ref{fig:koch_variations}, assuming $\tau_\mathrm{central}$ = 40.
(b) and (d): Relative differences in [\%] of observed MODIS radiances for wavelength channels 1, 5, 6, and 7 (assuming central wavelengths of 645, 1240, 1640, and 2130~nm) associated respectively to the extinction field modifications in (a) and (c).
For MODIS' (whisk-broom) camera, the mean VZA of 9$^\circ$ was assumed, averaged over the whole MISR swath inside of MODIS' much larger one.
Furthermore, the MYSTIC results were averaged and interpolated to a spatial resolution of 250~m for channel 1 and 500~m for the remaining channels.
}
\label{fig:koch_variations_modis}
\end{figure*}
The RT simulations in the previous section were performed for MISR's red band at 670~nm.
Light interaction with water droplets at this wavelength is dominated by scattering and absorption is negligible.
Figure~\ref{fig:koch_variations_modis} shows the relative differences of the radiances simulated with MYSTIC using the same clouds as in Fig.~\ref{fig:koch_variations} but for MODIS.
This instrument is a single downward-looking ``whisk-broom'' imager, but it provides invaluable additional information about cloud microphysics from its SWIR channels.
Here, we assume a mean VZA = 9$^\circ$, averaged over the $\approx$400~km MISR swath near the center of MODIS' 2330~km counterpart.
For MODIS' SWIR channels at 1240, 1640, and 2130~nm, absorption by water droplets becomes increasingly more important, which is exploited by the operational Nakajima--King retrieval \citep{nakajima90} to gain information about the effective droplet radius.

To investigate the effect of additional absorption on the veiled core of convective clouds, RT simulations were performed for MODIS channels 1, 5, 6, and 7, assuming central wavelengths of 645, 1240, 1640, and 2130~nm, which are indicated in Fig.~\ref{fig:koch_variations_modis}b and \ref{fig:koch_variations_modis}d in blue, green, orange and red, respectively.
In a post-processing step, the MYSTIC-computed radiances at the spatial resolution of about 4.3~m (turbulence grid-scale) were averaged and interpolated to 250~m for channel 1 and 500~m for channels 5, 6, and 7.

The relative differences of the radiances $\Delta I/I$ are well below 5\% for all MODIS pixels.
For increasing wavelengths from channel 1 to 7 the differences decrease, as expected, due to increasing absorption by the cloud droplets.
Observing the cloud scene at an absorbing wavelength leads to shorter photon paths inside the cloud, which in turn causes the veiled core to extend towards a smaller optical distance.

\section{Conclusions \& Outlook}
\label{sec:concl}

In this study, we investigated the ``veiled core'' inside optically thick convective clouds, as observed with MISR and MODIS on Terra.
We addressed two questions:
first, is there a veiled core in opaque convective clouds and, second, where is it?
To answer these questions, we simulated multi-angle observations of MISR using simplified 2D cloud models with a Koch fractal outer shape and high-resolution internal turbulence.
The results presented in this study indicate that a veiled core exists at an optical distance of $\approx$5 or more inside the cloud.
Changes in the liquid water distribution inside this core result in less than 5\% variability of the the multi-angle radiances observed by MISR, as long as mean, variance, and correlations of the liquid water field are preserved.
For MISR and MODIS, with an estimated measurement uncertainty of about 3\%, a variability of 5\% or less can still be considered within the noise, and is thus negligible in the sense that it should not be viewed as signal in cloud tomography.

The computational methodology used in this paper is complemented in a companion paper by \cite{Davis_etal19} where the physics of core veiling are explained in detail, and ensuing insights support the development of satellite cloud tomography.

The results from this 2D study can be directly transferred to 3D clouds.
In general, making use of this veiled core opens up new ways to increase the efficiency of rendering synthetic observations of optically thick cloud volumes using RT models.
The RT solver can be optimized to treat the outer shell of the cloud (where details matter) with higher accuracy than the veiled core, where the details of the liquid water distribution are negligible within the instrument's measurement uncertainty.
Applying photon diffusion approximation inside the veiled core is a promising way to significantly increase the speed and efficiency for RT simulations of optically thick clouds.
Replacing the detailed RT solver by a fast and efficient way to stream radiation through the veiled core, a significant amount of computational time and computer memory resources can be saved.
Since the size of the veiled core increases with a larger cloud volume and higher optical thickness, the increase in computational effort will be balanced.

For application specifically to the 3D cloud tomography inversion problem, making use of the veiled core is a promising way to increase the efficiency of the forward RT solver.
During the reconstruction of the cloud volume the forward solver has to be called at each iteration.
Thus, to speed up the reconstruction algorithm it is most important to increase the efficiency of the expensive RT solver.
Moreover, treating the veiled core as a connected object, mainly determined by its bulk optical properties significantly reduces the number of unknowns in the inverse problem.
Instead of solving the minimization for each pixel separately, only the bulk statistical properties of the core volume have to be estimated.
The statistical properties comprise mean, variance and correlation of the extinction volume, which can be parameterized for the RT solver.

Apart from 3D cloud tomography, other applications that involve RT modeling of optically thick cloud volumes can benefit from acknowledging the existence of a veiled core.
For example, the efficiency of RT modeling for the analysis of ground-based cloud images (e.g., at DOE/ARM sites) as demonstrated by \cite{romps2018} could be increased by making use of the veiled core.
Another application is the estimate of surface radiation budget at satellite pixel scales, e.g., for near-real time support of solar energy harvesting.
Here, fast RT modeling of cloudy scenes is needed for characterizing broken cloud structure which can help to mitigate the current fluctuations they induce to the electrical grid \citep{mejia2018}.
Moreover, satellite remote sensing of aerosols in the vicinity of clouds can benefit from this approach.
In these regions, we need to characterize the optical properties of aerosol and cloud particles most accurately to unravel indirect aerosol effects.
At the same time 1D RT-based aerosol retrievals fail here due to 3D cloud radiative effects \citep{varnai2017}.

%








%

\appendix[A]
\appendixtitle{MISR's Perspective on the Veiled Core: Sensitivity to SZA}
\label{sec:appendixA}

\begin{figure*}[t]
\includegraphics[width=\textwidth]{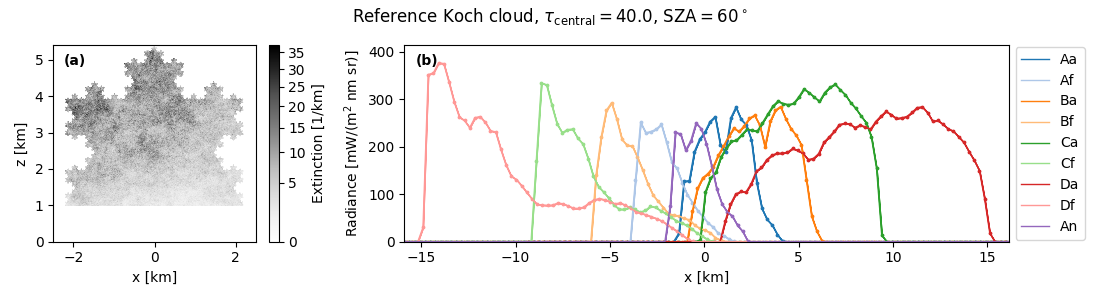}
\caption{
(a) Extinction field of the reference Koch cloud with $\tau_\mathrm{central}$ = 40.
(b) MISR radiances simulated for its nine cameras as in Fig.~\ref{fig:koch_highres_orig}b, but for SZA = 60$^\circ$.
}
\label{fig:koch_orig_sza60}
\end{figure*}

\begin{figure*}[t]
\includegraphics[width=\textwidth]{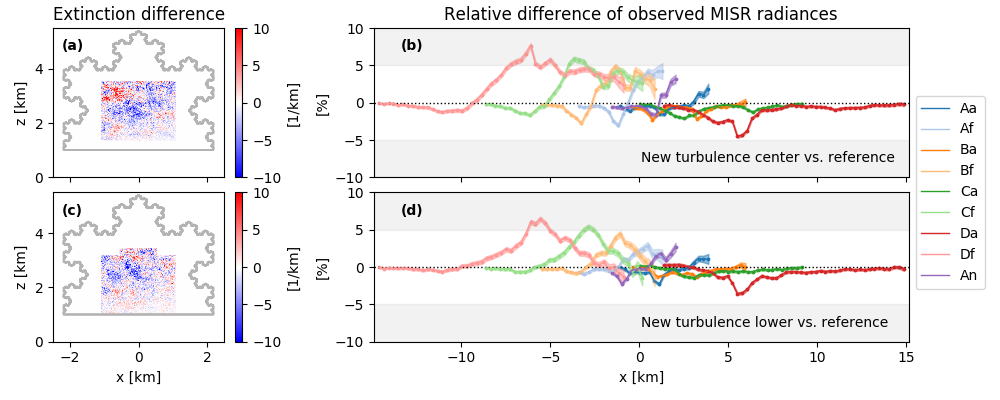}
\caption{
(a) and (c): Relative difference (new minus reference) of the extinction field, as in same panels for Fig.~\ref{fig:koch_variations}.
(b) and (d): Relative differences of observed MISR radiances as in same panels for Fig.~\ref{fig:koch_variations}, but for SZA = 60$^\circ$.
}
\label{fig:koch_variations_sza60}
\end{figure*}

Figure~\ref{fig:koch_orig_sza60} shows synthetic MISR observations from all nine cameras for the same Koch cloud as used in Fig.~\ref{fig:koch_highres_orig}, but assuming an oblique sun at SZA = 60$^\circ$, which is considered the average SZA, and the eight off-nadir viewing directions are in the principal plane.
Compared with MYSTIC simulation outcomes for overhead sun (SZA = 0$^\circ$), these MISR radiances here are clearly more asymmetric.
The sun is illuminating the cloud from the right, whereas its left side is in the shade.
This effect is visible in the observed MISR radiances which are now larger for the forward facing cameras (Af to Df) and decreased for the aftward facing cameras (Aa to Da) that are facing the self-shaded side of the cloud.

Figure~\ref{fig:koch_variations_sza60} shows the same sensitivity study as in Fig.~\ref{fig:koch_variations}, but for $\mathrm{SZA} = 60^\circ$.
In comparison with the results for overhead sun, the relative difference in the observed MISR radiances is more asymmetric as well.
The differences are larger for the forward facing cameras, which are pointing at the illuminated side of the cloud, which also happens to be the more tenuous side.
For this oblique illumination, photons thus travel a shorter optical distance through the cloud side, so there is less multiple scattering to ``veil'' the manipulated core.
Most of the $\Delta I/I$ values are still within the 5\%, which we still consider as the sensitivity threshold of MISR's sensors, and no new physics are required to explain the findings.

\appendix[B]
\appendixtitle{Idealized Convective Cloud Model}
\label{sec:appendixB}

Herein, we explain the model used to generate the idealized two-dimensional (2D) convective clouds used in this study.

The 2D synthetic clouds vary only in the $x$- and $z$-directions, and their geometric shape is derived from a Koch triangle \citep{koch1904}, which has an infinitely long perimeter with fractal dimension $\log 4/\log 3 \approx 1.26$ \citep{Mandelbrot77}, i.e., more than a polygon but less than a surface.  
This is how we approximate in a 2D world the multi-scale convective growth of clouds.
Using a 3D LES dynamical model, \cite{SiebesmaJonker00} have reproduced the observed \citep{Lovejoy1982} fractal structure of real-world cloud boundaries with fractal dimension $7/3 \approx 2.33$, i.e., more than a polyhedron but less than a volume.

The Koch fractal, which lives on a triangular grid, is mapped here to a rectangular grid for our present purposes, that is, to serve as input for the MYSTIC 3D RT solver.
The synthetic cloud is defined as an arbitrary (liquid water, extinction, or droplet density) field discretized on a grid of $1025\times1024$ (width $\times$ height) cells. 
As a physically relevant variation on the basic fractal model, a density decay factor of 0.9 is applied at each scale reduction.
Specifically, 2D density decreases multiplicatively by a factor of 0.9 for each fractal iteration, i.e., as smaller and smaller triangles are added to the structure of the fractal boundary of the Koch set.
That way, the cloud model does not have a ``hard'' boundary, even if it is fractal in nature.
Rather, it blends gradually into its clear-air environment, as is expected and observed in real-world clouds.

In addition, this outer fractal cloud structure with an ad hoc scale-by-scale decay is modulated internally with a judiciously chosen vertical gradient function, as follows.
``Rising parcel'' theory in cloud physics predicts strong vertical stratification of clouds following an adiabatic gradient in their convective cores, which is amply confirmed by observations.
Specifically, this gradient follows a linear trend in LWC over the vertical extent of the cloud, reaching its maximum typically in the upper third of the cloud from cloud top (e.g., \cite{pawlowska2000}).
LWC is converted to cloud extinction $\sigma_\mathrm{e}$ via the large-particle approximation in Mie scattering theory:
\begin{equation}
\sigma_\mathrm{e} \approx \frac{3}{2} \, \frac{\mathrm{LWC}}{r_\mathrm{eff} \, \rho_\mathrm{w}} \, ,
\label{eq:tau}
\end{equation}
where density of liquid water $\rho_\mathrm{w}$ = 10$^3$~kg/m$^3$.
Diffusional droplet growth predicts a slow 1/3 power law increase in $r_\mathrm{eff}$ with height above cloud base, assuming a constant droplet number concentration (\cite{Grosvenor_etal2018}, and references therein).
The adiabatic linear trend in LWC then leads to a 2/3 power law in extinction, from (\ref{eq:tau}).

\begin{figure}[t]
\includegraphics[width=8.3cm]{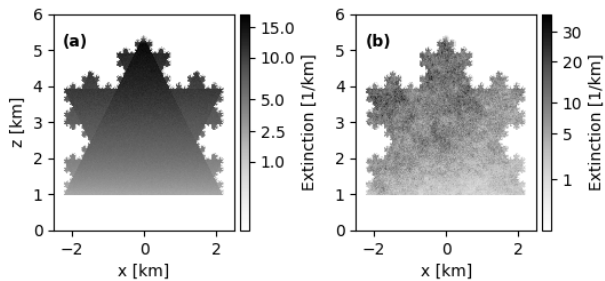}
\caption{
(a) Extinction field of Koch cloud model showing the fractal geometric shape with a slowly decaying extinction value for each iteration (i.e., smaller and smaller triangles) overlaid by a vertical gradient prescribed by an approximation for diffusional droplet growth, as explained in the main text.
(b) The extinction field in (a) modulated by a fBm-based turbulence field with Hurst exponent $H$ = 1/3.
Both extinction fields are scaled to a vertical optical thickness of $\tau_\mathrm{center}$ = 40 at cloud center ($x$ = 0).
}
\label{fig:koch_clouds}
\end{figure}

For simplicity, we follow \cite{Davis2008} and approximate (in the least-squares sense), over a finite range $0 \le z \le L_z$, any slow power-law increase $(z/L_z)^\gamma$ with an exponent $\gamma < 1$ with a linear function and a zero-level offset.  
The defining min-to-max ratio of this linear model is
\begin{equation*}
R_\gamma = \frac{1-\gamma}{1+2\gamma}.
\end{equation*}
For $\gamma = 2/3$, this zero-level offset is therefore at 1/7$^\text{th}$ of the maximum value, which we can take as unity without loss of generality.  
For added realism, we set the maximum to be reached at $z_\text{max}$ located at 1/10$^\text{th}$ of the cloud's thickness $L_z$ from its top, and reverse the linear trend back to the zero-level offset reached at cloud top ($z = L_z$).
In summary, the vertical gradient modulation function is therefore
\begin{equation*}
\max\left[ R_\gamma + (1-R_\gamma)\,\frac{z}{z_\text{max}}, 1 - (1-R_\gamma)\,\frac{z-z_\text{max}}{L_z-z_\text{max}} \right],
\end{equation*}
letting $z$ be the difference in altitude with cloud base (at $z$ = 0).  The resulting extinction field, which is scaled to a central optical thickness of $\tau_\mathrm{center} = 40$, is displayed in Fig.~\ref{fig:koch_clouds}a.

To simulate the convective cloud's internal structure, the extinction field in Fig.~\ref{fig:koch_clouds}a is modulated multiplicatively by an additional ``turbulence'' field.
The turbulence is modeled using 2D fBm with a Hurst exponent of $H = 1/3$, which is associated with a wavenumber spectral slope of $\beta = 2H + 1 = 5/3$.
The Hurst exponent takes values between $H \in [0, 1]$, and is a measure of long-range spatial correlations in turbulence.
For $H = 0.5$, the process corresponds to a 2D version of standard Brownian motion, with $\beta = 2$, where successive increments (e.g., actual steps in a random walk unfolding in time) are independent in magnitude and in sign.
For $H > 0.5$ the increments of the process exhibit a long-term positive autocorrelation, with $\beta > 2$, ending with smooth (at least once-differentiable) fields with $H = 1$ and $\beta \ge 3$.
This means that a high value in the series will probably be followed by another high value.
It also implies that the values a long time into the future will also tend to be high.
A Hurst exponent of $H < 0.5$, hence $1 \le \beta < 2$, which is the case in this study, corresponds to successive increments of the process being negatively correlated. 
This means that high and low values tend to alternate.

The fBm is modeled using the diamond-square algorithm, a method originally used for generating terrain height maps for computer graphics, first introduced by \citet{fournier1982}.
It starts with a 2D grid of width and height $2^N+1$, where we choose to set $N$ = 10.
This choice ensures that the turbulence field has scaling spatial statistics (e.g., its wavenumber spectrum) that covers 3 orders-of-magnitude from the cloud scale to the grid-cell scale, specifically, from a few km to a few m, as is observed in nature (\cite{Davis_etal99}, and references therein).
After initializing the four corner values, the diamond and square steps are performed alternately at smaller and smaller scales until all values of the field are determined; see illustration at (https://en.wikipedia.org/wiki/Diamond-square\_algorithm).
To avoid large random gradients at the scale of the whole cloud, we initialized the four corners of the domain to null values.
For the diamond step, the midpoint of each square in the array is computed by the average of the four corner points, plus a random (zero-mean, normal) value.
For the square step, the midpoint of each diamond is set to the average of the four corner points (only three if it is at the boundary), plus a random value.
At each change in scale, the magnitude (standard deviation) of the random (normal) variables is reduced by a factor 1/2$^H \approx$ 0.794 when $H$ = 1/3.
The final cloud field, including turbulence, is tuned to have a realistic log-normal one-point probability density function with unit mean and unit variance.

One drawback of this method is that it produces slight vertical and horizontal ``creases'' since the most significant perturbations take place in a rectangular grid.
More sophisticated (Fourier-space) methods have been developed to compute 2D fBm fields (e.g., \citet{stein2002}, \citet{kroese2015}) that overcome this issue.
However, the diamond-square algorithm has an important advantage for the present study.
This method allows different randomly generated turbulence fields to be stitched together without discontinuities simply by disabling the addition of a random increment at all boundary points.
This advantage was exploited here to replace a quarter of the original turbulence field or more---and an even larger fraction of the 2D Koch cloud volume---by a new realization of the fBm field, as is shown in Fig.~\ref{fig:koch_variations}a and c.

Finally, the central optical thickness is computed by integrating the cloud extinction along the vertical axis at $x$ = 0:
\begin{equation}
\tau_\mathrm{center} = \sum_{i=1}^{1024} \sigma_\text{e}(i,j)\, \Delta z \, ,
\end{equation}
where $\Delta z = L_z/1024$ is the vertical grid-scale, after the whole extinction grid $\sigma_\text{e}(i,j)$ is scaled to yield a desired value of $\tau_\mathrm{center}$.




%
%

%
\vspace{24pt}
\acknowledgments
This work was performed at the Jet Propulsion Laboratory, California Institute of Technology under contract with the National Aeronautics and Space Administration.
LF received funding from the European Union's Framework Programme for Research and Innovation Horizon 2020 (2014-2020) under the Marie Sk\l{}odowska-Curie Grant Agreement No. 754388 and from LMU Munich's Institutional Strategy LMUexcellent within the framework of the German Excellence Initiative (No. ZUK22).
AD was funded by by NASA's SMD/ESD Radiation Sciences Program under the ROSES TASNPP element (contract \#17-TASNPP17-0165).
We acknowledge many fruitful conversations on cloud CT with Aviad Levis, Masada Tzemach, Yoav Schechner, Alex B. Kostinski, Jesse Loveridge, Larry Di Girolamo, and Katie Bouman.
We thank Prof. Yuk L. Yung, Division of Geological and Planetary Sciences, California Institute of Technology, for his enthusiastic support for this project.

\bibliographystyle{ametsoc2014}

\begin{thebibliography}{41}
\providecommand{\natexlab}[1]{#1}
\providecommand{\url}[1]{\texttt{#1}}
\renewcommand{\UrlFont}{\rmfamily}
\providecommand{\urlprefix}{URL }
\expandafter\ifx\csname urlstyle\endcsname\relax
  \providecommand{\doi}[1]{doi:\discretionary{}{}{}#1}\else
  \providecommand{\doi}{doi:\discretionary{}{}{}\begingroup
  \urlstyle{rm}\Url}\fi
\providecommand{\eprint}[2][]{\url{#2}}

\bibitem[{{Barnes} et~al.(1998){Barnes}, {Pagano},, and
  {Salomonson}}]{barnes1998}
{Barnes}, W.~L., T.~S. {Pagano}, and V.~V. {Salomonson}, 1998: Prelaunch
  characteristics of the {M}oderate {R}esolution {I}maging {S}pectroradiometer
  ({MODIS}) on {EOS-AM1}. \textit{IEEE Transactions on Geoscience and Remote
  Sensing}, \textbf{36~(4)}, 1088--1100, \doi{10.1109/36.700993}.

\bibitem[{Bony et~al.(2017)}]{bony2017}
Bony, S., and Coauthors, 2017: {{EUREC4A: A} Field Campaign to {E}lucidate the
  {C}ouplings Between {C}louds, {C}onvection and {C}irculation}. \textit{Surveys in
  Geophysics}, \doi{10.1007/s10712-017-9428-0}.

\bibitem[{Boucher et~al.(2013)}]{boucher2013}
Boucher, O., and Coauthors, 2013: Clouds and aerosols. \textit{{Climate Change
  2013: The Physical Science Basis. Contribution of Working Group I to the
  Fifth Assessment Report of the Intergovernmental Panel on Climate Change}},
  T.~F. Stocker, D.~Qin, G.-K. Plattner, M.~Tignor, S.~Allen, J.~Boschung,
  A.~Nauels, Y.~Xia, V.~Bex, and P.~Midgley, Eds., Cambridge.

\bibitem[{Buras et~al.(2011)Buras, Dowling,, and Emde}]{buras2011}
Buras, R., T.~Dowling, and C.~Emde, 2011: {New secondary-scattering correction
  in {DISORT} with increased efficiency for forward scattering}. \textit{{J.
  Quant. Spectrosc. Radiat. Transfer}}, \textbf{112~(12)}, 2028--2034,
  \doi{10.1016/j.jqsrt.2011.03.019}.

\bibitem[{Cho et~al.(2015)}]{cho2015}
Cho, H.-M., and Coauthors, 2015: Frequency and causes of failed {MODIS} cloud
  property retrievals for liquid phase clouds over global oceans.
  \textit{Journal of Geophysical Research: Atmospheres}, \textbf{120~(9)},
  4132--4154, \doi{10.1002/2015JD023161}.

\bibitem[{Cornet and Davies(2008)Cornet, and Davies}]{cornet2008}
Cornet, C., and R.~Davies, 2008: {Use of {MISR} measurements to study the
  radiative transfer of an isolated convective cloud: {I}mplications for cloud
  optical thickness retrieval}. \textit{Journal of Geophysical Research:
  Atmospheres}, \textbf{113~(D4)}, \doi{10.1029/2007JD008921}.

\bibitem[{Davis(2008)}]{Davis2008}
Davis, A.~B., 2008: Multiple-scattering lidar from both sides of the clouds:
  {A}ddressing internal structure. \textit{J. Geophys. Res.}, \textbf{113},
  D14S10, \doi{10.1029/2007JD009666}.

\bibitem[{Davis et~al.(2019)Davis, Forster, Diner,, and Mayer}]{Davis_etal19}
Davis, A.~B., L.~Forster, D.~J. Diner, and B.~Mayer, 2019: Cloud tomography
  from space using {MISR} and {MODIS}: The physics of ``core veiling'' in opaque
  convective clouds. \textit{J. Atmos. Sci.}, (in preparation).

\bibitem[{Davis and Marshak(2004)Davis, and Marshak}]{DavisMarshak04}
Davis, A.~B., and A.~Marshak, 2004: Photon propagation in heterogeneous optical
  media with spatial correlations: {E}nhanced mean-free-paths and
  wider-than-exponential free-path distributions. \textit{J. Quant. Spectrosc.
  Rad. Transf.}, \textbf{84}, 3--34.

\bibitem[{Davis et~al.(1999)Davis, Marshak, Gerber,, and
  Wiscombe}]{Davis_etal99}
Davis, A.~B., A.~Marshak, H.~Gerber, and W.~J.~Wiscombe, 1999: Horizontal
  structure of marine boundary-layer clouds from cm- to km-scales. \textit{J.
  Geophys. Res.}, \textbf{D104}, 6123--6144.

\bibitem[{{Diner} et~al.(1998)}]{diner1998}
{Diner}, D.~J., and Coauthors, 1998: {Multi-angle Imaging SpectroRadiometer
  (MISR) instrument description and experiment overview}. \textit{IEEE
  Transactions on Geoscience and Remote Sensing}, \textbf{36}, 1072--1087,
  \doi{10.1109/36.700992}.

\bibitem[{Diner et~al.(2013)}]{diner2013}
Diner, D.~J., and Coauthors, 2013: {The Airborne Multiangle SpectroPolarimetric
  Imager (AirMSPI): {A} new tool for aerosol and cloud remote sensing}.
  \textit{{Atmos. Meas. Tech.}}, \textbf{6~(8)}, 2007--2025,
  \doi{10.5194/amt-6-2007-2013}.

\bibitem[{Emde et~al.(2016)}]{emde2016}
Emde, C., and Coauthors, 2016: {The libRadtran software package for radiative
  transfer calculations (version 2.0.1)}. \textit{Geosci. Model Dev.},
  \textbf{9~(5)}, 1647--1672, \doi{10.5194/gmd-9-1647-2016}.

\bibitem[{Evans(1998)}]{evans98}
Evans, K.~F., 1998: {The spherical harmonics discrete ordinate method for
  three-dimensional atmospheric radiative transfer}. \textit{{J. Atmos. Sci.}},
  \textbf{55}, 429--446, \doi{10.1175/1520-0469(1998)055<0429:TSHDOM>2.0.CO;2}.

\bibitem[{Fournier et~al.(1982)Fournier, Fussell,, and
  Carpenter}]{fournier1982}
Fournier, A., D.~Fussell, and L.~Carpenter, 1982: Computer rendering of
  stochastic models. \textit{Commun. ACM}, \textbf{25~(6)}, 371--384,
  \doi{10.1145/358523.358553}.

\bibitem[{Grosvenor et~al.(2018)}]{Grosvenor_etal2018}
Grosvenor, D.~P., and Coauthors, 2018: Remote sensing of droplet number
  concentration in warm clouds: {A} review of the current state of knowledge and
  perspectives. \textit{Reviews of Geophysics}, \textbf{56~(2)}, 409--453,
  \doi{10.1029/2017RG000593}.

\bibitem[{Jensen et~al.(2016)}]{jensen2016}
Jensen, M.~P., and Coauthors, 2016: The {M}idlatitude {C}ontinental
  {C}onvective {C}louds {E}xperiment ({MC3E}). \textit{{Bull. Amer. Meteor.
  Soc.}}, \textbf{97~(9)}, 1667--1686, \doi{10.1175/BAMS-D-14-00228.1}.

\bibitem[{King et~al.(2003)}]{king2003}
King, M.~D., and Coauthors, 2003: {Cloud and aerosol properties, precipitable
  water, and profiles of temperature and water vapor from MODIS}. \textit{IEEE
  Transactions on Geoscience and Remote Sensing}, \textbf{41}, 442--458,
  \doi{10.1109/TGRS.2002.808226}.

\bibitem[{Kroese and Botev(2015)Kroese, and Botev}]{kroese2015}
Kroese, D.~P., and Z.~I. Botev, 2015: \textit{Spatial Process Simulation},
  369--404. Springer International Publishing, Cham,
  \doi{10.1007/978-3-319-10064-7_12}.

\bibitem[{Lee et~al.(2018)Lee, Di~Girolamo, Zhao,, and Zhan}]{lee2018}
Lee, B., L.~Di~Girolamo, G.~Zhao, and Y.~Zhan, 2018: Three-dimensional cloud
  volume reconstruction from the {M}ulti-angle {I}maging {S}pectro{R}adiometer.
  \textit{Remote Sensing}, \textbf{10~(11)}, 1858, \doi{10.3390/rs10111858},
  \urlprefix\url{http://dx.doi.org/10.3390/rs10111858}.

\bibitem[{{Levis} et~al.(2015){Levis}, {Schechner}, {Aides},, and
  {Davis}}]{levis2015}
{Levis}, A., Y.~Y. {Schechner}, A.~{Aides}, and A.~B. {Davis}, 2015: Airborne
  three-dimensional cloud tomography. \textit{2015 IEEE International
  Conference on Computer Vision (ICCV)}, 3379--3387,
  \doi{10.1109/ICCV.2015.386}.

\bibitem[{{Levis} et~al.(2017){Levis}, {Schechner},, and {Davis}}]{levis2017}
{Levis}, A., Y.~Y. {Schechner}, and A.~B. {Davis}, 2017: Multiple-scattering
  microphysics tomography. \textit{2017 IEEE Conference on Computer Vision and
  Pattern Recognition (CVPR)}, 5797--5806, \doi{10.1109/CVPR.2017.614}.

\bibitem[{Lovejoy(1982)}]{Lovejoy1982}
Lovejoy, S., 1982: {Area-Perimeter Relation for Rain and Cloud Areas}.
  \textit{Science}, \textbf{216~(4542)}, 185--187,
  \doi{10.1126/science.216.4542.185}.

\bibitem[{Mandelbrot(1977)}]{Mandelbrot77}
Mandelbrot, B.~B., 1977: \textit{Fractals: Form, Chance, and Dimension}. W.H.
  Freeman, San Francisco (Ca).

\bibitem[{Matheou and Chung(2014)Matheou, and Chung}]{MatheouChung14}
Matheou, G., and D.~Chung, 2014: Large-eddy simulation of stratified
  turbulence. {P}art {II}: {A}pplication of the stretched-vortex model to the
  atmospheric boundary layer. \textit{Journal of the Atmospheric Sciences},
  \textbf{71~(12)}, 4439--4460, \doi{10.1175/JAS-D-13-0306.1}.

\bibitem[{Mayer(2009)}]{mayer2009}
Mayer, B., 2009: {Radiative transfer in the cloudy atmosphere}.
  \textit{European Physical Journal Conferences}, \textbf{1}, 75--99,
  \doi{10.1140/epjconf/e2009-00912-1}.

\bibitem[{Mayer and Kylling(2005)Mayer, and Kylling}]{mayer2005}
Mayer, B., and A.~Kylling, 2005: {Technical Note: The libRadtran software
  package for radiative transfer calculations: Description and examples of
  use}. \textit{{Atmos. Chem. Phys.}}, \textbf{5}, 1855--1877,
  \doi{10.5194/acp-5-1855-2005}.

\bibitem[{Mejia et~al.(2018)Mejia, Kurtz, Levis, Íñigo de~la Parra,, and
  Kleissl}]{mejia2018}
Mejia, F.~A., B.~Kurtz, A.~Levis, Íñigo de~la Parra, and J.~Kleissl, 2018:
  Cloud tomography applied to sky images: {A} virtual testbed. \textit{Solar
  Energy}, \textbf{176}, 287 -- 300,
  \doi{https://doi.org/10.1016/j.solener.2018.10.023}.

\bibitem[{Mie(1908)}]{mie08}
Mie, G., 1908: {Beitr{\"a}ge zur Optik tr{\"u}ber Medien, speziell kolloidaler
  Metall{\"o}sungen}. \textit{Annalen der Physik, Vierte Folge},
  \textbf{25~(3)}, 377--445.

\bibitem[{Nakajima and King(1990)Nakajima, and King}]{nakajima90}
Nakajima, T., and M.~King, 1990: {Determination of the optical thickness and
  effective particle radius of clouds from reflected solar radiation
  measurements. Part I: Theory}. \textit{{J. Atmos. Sci.}}, \textbf{47},
  1878--1893, \doi{10.1175/1520-0469(1990)047<1878:DOTOTA>2.0.CO;2}.

\bibitem[{Pawlowska et~al.(2000)Pawlowska, Brenguier,, and
  Burnet}]{pawlowska2000}
Pawlowska, H., J.~Brenguier, and F.~Burnet, 2000: {Microphysical properties of
  stratocumulus clouds}. \textit{Atmospheric Research}, \textbf{55}, 15--33,
  \doi{https://doi.org/10.1016/S0169-8095(00)00054-5}.

\bibitem[{Platnick et~al.(2003)Platnick, King, Ackerman, Menzel, Baum, Riedi,,
  and Frey}]{platnick2003}
Platnick, S., M.~King, S.~Ackerman, W.~Menzel, B.~Baum, J.~Riedi, and R.~Frey,
  2003: {The MODIS cloud products: Algorithms and examples from TERRA}.
  \textit{IEEE Transactions on Geoscience and Remote Sensing}, \textbf{41},
  459--473, \doi{10.1109/tgrs.2002.808301}.

\bibitem[{Romps and {\"O}ktem(2018)Romps, and {\"O}ktem}]{romps2018}
Romps, D.~M., and R.~{\"O}ktem, 2018: Observing clouds in 4{D} with multiview
  stereophotogrammetry. \textit{Bulletin of the American Meteorological
  Society}, \textbf{99~(12)}, 2575--2586, \doi{10.1175/BAMS-D-18-0029.1}.

\bibitem[{Seiz and Davies(2006)Seiz, and Davies}]{seiz2006}
Seiz, G., and R.~Davies, 2006: {Reconstruction of cloud geometry from
  multi-view satellite images}. \textit{Remote Sensing of Environment},
  \textbf{100~(2)}, 143--149, \doi{10.1016/j.rse.2005.09.016}.

\bibitem[{Siebesma and Jonker(2000)Siebesma, and Jonker}]{SiebesmaJonker00}
Siebesma, A.~P., and H.~J.~J. Jonker, 2000: Anomalous scaling of cumulus cloud
  boundaries. \textit{Phys. Rev. Lett.}, \textbf{85}, 214--217.

\bibitem[{Stein(2002)}]{stein2002}
Stein, M.~L., 2002: Fast and exact simulation of fractional brownian surfaces.
  \textit{Journal of Computational and Graphical Statistics}, \textbf{11~(3)},
  587--599, \doi{10.1198/106186002466}.

\bibitem[{Thuillier et~al.(2003)Thuillier, Hers{\'e}, Labs, Foujols,
  Peetermans, Gillotay, Simon,, and Mandel}]{thuillier2003}
Thuillier, G., M.~Hers{\'e}, D.~Labs, T.~Foujols, W.~Peetermans, D.~Gillotay,
  P.~Simon, and H.~Mandel, 2003: The solar spectral irradiance from 200 to 2400
  nm as measured by the {SOLSPEC} spectrometer from the {A}tlas and {E}ureca
  missions. \textit{Solar Physics}, \textbf{214~(1)}, 1--22,
  \doi{10.1023/A:1024048429145}.

\bibitem[{{von Koch}(1904)}]{koch1904}
{von Koch}, H., 1904: {Sur une courbe continue sans tangente, obtenue par une
  construction g\'eom\'etrique \'el\'ementaire.} \textit{{Ark. Mat. Astron.
  Fys.}}, \textbf{1}, 681--702.

\bibitem[{V\'{a}rnai et~al.(2017)V\'{a}rnai, Marshak,, and Eck}]{varnai2017}
V\'{a}rnai, T., A.~Marshak, and T.~F. Eck, 2017: Observation-based study on
  aerosol optical depth and particle size in partly cloudy regions.
  \textit{Journal of Geophysical Research: Atmospheres}, \textbf{122~(18)},
  10,013--10,024, \doi{10.1002/2017JD027028}.

\bibitem[{Wiscombe(1980)}]{wiscombe80a}
Wiscombe, W.~J., 1980: {Improved Mie scattering algorithms}. \textit{{Applied
  Optics}}, \textbf{19~(9)}, 1505--1509, \doi{10.1364/AO.19.001505}.

\bibitem[{Zhang and Platnick(2011)Zhang, and Platnick}]{zhang2011}
Zhang, Z., and S.~Platnick, 2011: An assessment of differences between cloud
  effective particle radius retrievals for marine water clouds from three {MODIS}
  spectral bands. \textit{{Journal of Geophysical Research}},
  \textbf{116~(D20)}, \doi{10.1029/2011JD016216}.

\end{thebibliography}
\newcommand{\noop}[1]{}

\vspace{24pt}
\copyright 2019 California Institute of Technology. Government sponsorship acknowledged.

%

%

\end{document}